\documentstyle[12pt,cmb]{article}
\input epsf

\def\vec{\bf}

\def\hmpc{{\, {\rm h}^{-1}~\rm Mpc}}

\def\kpc{{\rm~kpc}}

\def\'{^{\prime}}
\def\avrg#1{{\langle #1 \rangle}}

\def\bv{{\vec v}}

\def\hq{{\hat q}}

\def\eps{\varepsilon}

\def\eg{{e.g., }}
\def\ie{{i.e., }}
\def\etal{{et al. }}
 
\def\etc{{etc. }}

\def\half{{\textstyle{1\over2}}}

\def\p3m{P$^3$M}

\def\spose#1{\hbox to 0pt{#1\hss}}
\def\lta{\mathrel{\spose{\lower 3pt\hbox{$\mathchar"218$}}
     \raise 2.0pt\hbox{$\mathchar"13C$}}}
\def\gta{\mathrel{\spose{\lower 3pt\hbox{$\mathchar"218$}}
     \raise 2.0pt\hbox{$\mathchar"13E$}}}
\def\ge{\mathrel{\spose{\lower 3pt\hbox{$-$}}
     \raise 2.0pt\hbox{$\mathchar"13E$}}}
\def\le{\mathrel{\spose{\lower 3pt\hbox{$-$}}
     \raise 2.0pt\hbox{$\mathchar"13C$}}}

\def\eqright{\begin{eqnarray}}
\def\endeqright{\end{eqnarray}}

\begin{document}

\heading{THE THEORETICAL AGENDA IN CMB RESEARCH}


\author{J. Richard
Bond} {Canadian Institute
for Advanced Research Cosmology Program}{Canadian Institute
for Theoretical Astrophysics, Toronto, Ontario, Canada.} 

\begin{abstract}{\baselineskip 0.0cm 
\small The terrain that theorists cover in this CMB golden age is
described.  We ponder early universe physics in quest of the
fluctuation generator. We extoll the virtues of inflation and
defects. We transport fields, matter and radiation into the linear
(primary anisotropies) and nonlinear (secondary anisotropies)
regimes. We validate our linear codes to deliver accurate predictions
for experimentalists to shoot at. We struggle at the computing edge to
push our nonlinear simulations from only illustrative to fully
predictive. We are now phenomenologists, optimizing statistical
techniques for extracting truths and their errors from current and
future experiments. We begin to clean foregrounds. We join CMB
experimental teams. We combine the CMB with large scale structure,
galaxy and other cosmological observations in search of current
concordance. The brave use all topical data. Others carefully craft
their prior probabilities to downweight data sets. We are always
unbiased. We declare theories sick, dead, ugly. Sometimes we cure
them,  resurrect them, rarely beautify them. Our goal is to
understand how all cosmic structure we see arose and what the Universe
is made of, and to use this to discover the laws of ultrahigh energy
physics. Theorists are humble, without hubris.}
\end{abstract}

\section{Theoretical Theory}\label{sec:theory}

{\bf Early Universe Physics} is the likely source of fluctuations to
input into the cosmic structure formation problem (\eg
\cite{bh95}). We want to measure the CMB (and large scale structure)
response to these initial fluctuations. The goal is the lofty one of
peering into the physical mechanism by which the fluctuations were
generated, by learning: the statistics of the fluctuations, whether
Gaussian or non-Gaussian; the mode, whether adiabatic or isocurvature
scalar perturbations, and whether there is a significant component in
gravitational wave tensor perturbations; the power spectra for these
modes (albeit over only 3 decades in wavenumber for CMB probes,
Fig.~\ref{fig:probes}, and fewer for large scale structure (LSS)
probes).

{\bf Fluctuation Generation Mechanisms:} The contenders are (1) ``zero
point'' oscillations in scalar and tensor fields that must be there in
the early universe if quantum mechanics is applicable and (2)
topological defects which may arise in the inevitable phase
transitions expected in the early universe.

{\bf Inflation:} For quantum oscillations to be important, a period of
accelerated expansion seems essential, for then the comoving Hubble
length $(Ha)^{-1}$ shrinks in size, freezing the time-incoherent noise
into time-coherent patterns for structure formation.\footnote{Acceleration (\ie
inflation) seems to be generic unless it is explicitly forbidden by
some law of physics unknown at this time. It may be extremely
improbable {\it a priori} that a bit of space would accelerate and yet
be highly probable {\it a posteriori} that we would find ourselves in
a region that once inflated because of the vast space created.} A major
challenge for inflation models is a ``natural'' explanation of why
post-inflation fluctuations should be $\sim 10^{-5}$ in size, as
revealed by COBE and also LSS observations. 

 Many many variants of the basic inflation theme have been proposed,
sometimes with radically different consequences for the appearance of
the CMB sky, which is used in fact to highly constrain the more
baroque models.  A rank-ordering of inflation possibilities: (1)
adiabatic curvature fluctuations with nearly uniform scalar tilt over
the observable range, slightly more power to large scales ($n_s<1$)
than ``scale invariance'' ($n_s=1$) gives, a predictable nonzero
gravity wave contribution with tilt similar to the scalar one, and
tiny mean curvature ($\Omega_{tot}\approx 1$); (2) same as (1), but
with a tiny gravity wave contribution; (3) same as (1) but with a
subdominant isocurvature component of nearly scale invariant tilt (the
case in which isocurvature dominates is ruled out by $\Delta T/T$);
(4) radically broken scale invariance with weak to moderate features
(ramps, mountains, valleys) in the fluctuation spectrum (strong ones
are largely ruled out by $\Delta T/T$); (5) radical breaking with
non-Gaussian features as well; (6) ``open'' inflation, with quantum
tunneling producing a negatively-curved (hyperbolic) space which
inflates, but not so much as to flatten the mean curvature ($d_c \sim
(Ha)^{-1}$, not $\gg (Ha)^{-1}$, where $d_c\equiv H_0^{-1}\vert
1-\Omega_{tot} \vert^{-1/2}$); (7) quantum creation of compact
hyperbolic space from ``nothing'' with volume $d_T^3$ which inflates,
with $d_T \sim (Ha)^{-1}$, not $\gg (Ha)^{-1}$, and $d_T$ of order
$d_c$; (8) flat ($d_c=\infty $) inflating models which are small tori
of scale $d_T$ with $d_T$ a few $(Ha)^{-1}$ in size. It is quite
debatable which of the cases beyond (2) are more or less plausible,
with some claims that (4) is supersymmetry-inspired, others that (6)
is not as improbable as it sounds (see Cohn, these proceedings, for a
nice discussion). It is the theorists' job to push out the boundaries
of the inflation idea and use the data to select what is allowed.

{\bf Defects:} Gradients in the disordered field energy left by a
phase transition are smoothed on the growing scale over which causal
communication can occur. Topological knot-like or string-like field
configurations disappear slowly enough that they can act as
isocurvature seed perturbations to drive the growth of fluctuations in
the total mass density. The inherent nonlinearity of these defects in
the field implies many issues can only be answered with numerical
simulations, and these are subject to the same
computational-size-to-resolution limitations that plague the rest of
nonlinear cosmology.  Much analytic progress in understanding the
basic observable features of defect models is being made and renewed
effort is being put into doing large enough simulations to predict CMB
anisotropies in both texture and string theories.

{\bf Hydro:} Although hydrodynamic and radiative processes are
expected to play important roles around collapsed objects and may bias
the galaxy distribution relative to the mass, a global role in
obscuring the early universe fluctuations by late time generation on
large scales now seems unlikely.\footnote{Not too long ago it seemed
perfectly reasonable based on extrapolation from the physics of the
interstellar medium to the pregalactic and intergalactic medium to
suppose hydrodynamical amplification of seed cosmic structure would
obscure primordial fluctuations from the early Universe. The strong
limits on Compton cooling from FIRAS, in energy ${\delta E_{Compton\
cool} / E_{cmb}} = 4y < 6.0 \times 10^{-5}$ (95\% CL), constrain the
product $f_{exp}R_{exp}^2$ of filling factor $f_{exp}$ and bubble
formation scale $R_{exp}$, to values too small for a purely
hydrodynamic origin. If supernovae were responsible for the blasts,
the accompanying presupernova light radiated would have been much in
excess of the explosive energy (more than a hundred-fold), leading to
much stronger restrictions.}

{\bf Transport:} Cosmological radiative transfer is on a firm
theoretical footing. Together with a gravity
theory\footnote{Einstein's theory is invariably assumed, but
deviations are expected and indeed necessary at very high energy, with
potential impact on the fluctuation generation process, and, if exotic
enough, on transport through decoupling to now. Eventually, as we
understand the CMB sky better, the data will undoubtedly be turned to
constraining or discovering modified theories of gravity.} and the
transport theory for the other fields and particles present (baryons,
hot, warm and cold dark matter, coherent fields, \ie ``dynamical''
cosmological ``constants'', {\it etc.}), we propagate initial
fluctuations from the early universe through photon decoupling into the
(very) weakly nonlinear phase, and predict {\it primary anisotropies},
those calculated using either linear perturbation theory (\eg for
inflation-generated fluctuations), or, in the case of defects, linear
response theory. The sources driving their development are listed in
the Table. 

{\bf Cosmic Parameters:} Even simple Gaussian inflation-generated
fluctuations for structure formation have a large number of early
universe parameters we would wish to determine: power spectrum
amplitudes at some normalization wavenumber $k_n$ for the modes
present, $\{ {\cal P}_{ad}(k_n), {\cal P}_{is}(k_n) , {\cal
P}_{GW}(k_n) \}$; ``tilt'' shape functions $\{ n_s(k),n_{is}(k),
n_t(k) \} $, usually chosen to be constant or with a logarithmic
correction, \eg $n_s(k_n), dn_s(k_n)/d\ln k$. The transport problem is
dependent upon physical processes, and hence on physical parameters. A
partial list includes various mean energy densities $\{ \Omega_{tot},
\Omega_B , \Omega_{vac}, \Omega_{cdm }, \Omega_{hdm }\}$, the Hubble
parameter ${\rm h}$, the number of relativistic neutrinos, the
abundance of primordial helium, and parameters characterizing the
ionization history of the Universe, \eg the Compton optical depth from
a reheating redshift $z_{reh}$ to the present. For a given model, the
early universe power spectrum amplitude measures are uniquely related
to late-time power spectrum measures of relevance for the CMB, such as
the quadrupole or bandpowers for various experiments, or to large
scale structure observations, such as the {\it rms} density
fluctuation level on the $8\hmpc$ (cluster) scale, $\sigma_8$.

The arena in which theory battles observation is the anisotropy power
spectrum figure. Fig.~\ref{fig:CLpow} illustrates how primary ${\cal
C}_\ell$'s vary with cosmic parameters. They are normalized to the
4-year {\it dmr}(53+90+31)(A+B) data. A ``standard'' CDM model, with
$n_s$=1, $\Omega_{tot}$=1, $H_0$=50, $\Omega_B$=0.05, and a 13 Gyr
age, is the upper solid curve.  It has $\sigma_8=1.20\pm 0.08$, far
from the $\sim 0.6$ target value derived from cluster abundance
observations. An (almost indistinguishable) dotted curve has the same
parameters except that it includes a light neutrino with
$\Omega_{m\nu} =0.2$ (and $\Omega_{cdm}=0.75$). It has $\sigma_8
=0.83\pm 0.06$. The upper dashed curve is a vacuum-dominated model
with $H_0$=75 and the 13 Gyr age (and $\Omega_\Lambda =0.73$,
$\Omega_B=0.02$, $\Omega_{cdm}=0.24$). It has $\sigma_8=1.03\pm 0.07$,
which is OK for cluster abundances. An open CDM model has the ${\cal
C}_\ell$ peak shifted to higher $\ell$; the one shown has $H_0=60$ and
the 13 Gyr age, with $\Omega_{tot}$=0.33, $\Omega_{cdm}$=0.30,
$\Omega_B=0.035$, and $\sigma_8=0.50\pm 0.04$.  By $H_0=70$,
$\Omega_{tot}$ is down to 0.055 at this age. The lower solid curve is
a CDM model with reionization at $z_{reh}=30$, and almost degenerate
with it is a tilted CDM model ($n_s$=0.95 but otherwise
standard). Even the nearly degenerate hot/cold and CDM models shown
should be distinguishable by COBRAS/SAMBA.


{\bf COMBA:} Spurred on by the promise of percent-level precision in
cosmic parameters from CMB satellites, a considerable fraction of the
CMB theoretical community with Boltzmann transport codes compared
their approaches and validated the results to ensure percent-level
accuracy up to $\ell \sim 3000$ \cite{comba95}.  The arena shifted
from figures like Fig.~\ref{fig:CLpow} to $\Delta {\cal C}_\ell/{\cal
C}_\ell$ figures with tiny vertical range. We look forward to the
happy day when such a relative difference figure will be used to
reveal the remaining tiny residuals in the best fit theoretical
model. An important goal for COMBA was speed, since the parameter
space we wish to constrain has many dimensions.  Most groups have
solved cosmological radiative transport by evolving a hierarchy of
coupled moment equations, one for each $\ell$. Although the equations
and techniques were in place prior to the COBE discovery for scalar
modes, and shortly after for tensor modes, to get the high accuracy
with speed has been somewhat of a challenge. There are alternatives to
the moment hierarchy for the transport of photons and neutrinos. In
particular the entire problem of photon transport reduces to integral
equations in which the multipoles with $\ell >2$ are expressed as
history-integrals of metric variables, photon-bunching, Doppler and
polarization sources, as in the Table. 
The fastest COMBA-validated code uses this method (Seljak, these
proceedings).

{\bf Secondary:} Secondary anisotropies, with sources listed in
the Table,
arise from nonlinear structures. They are a
nuisance foreground to be subtracted to get at the {\it primary}
primary ones, but also invaluable probes of shorter-distance aspects
of structure formation theories, full of important cosmological
information. The effect of lensing is to smooth slightly the Doppler
peaks and troughs of Fig.~\ref{fig:CLpow}. ${\cal C}_\ell$'s from
quadratic nonlinearities in the gas at high redshift are concentrated
at high $\ell$, but for most viable models are expected to be a small
contaminant. Scattering from gas in moving clusters also has a small
effect on ${\cal C}_\ell$, although is measurable in individual
clusters.  Power spectra for the thermal SZ effect from clusters are
larger; examples in Fig.~\ref{fig:CLpow} are for a cluster-normalized
$\sigma_8$=0.7 hot/cold hybrid model (solid, $\Omega_{hdm} $=0.3) and
an $n_s$=0.8 tilted CDM model. Although ${\cal C}^{(SZ)}_\ell$ may be
small, because the power for such non-Gaussian sources is concentrated
in hot or cold spots the signal is detectable, and often has been.
${\cal C}_\ell$ for a typical dusty primeval galaxy model is also
shown, the larger (arbitrarily normalized) part a shot-noise effect
for galaxies with dust distributed over $10 \kpc$, the smaller a
contribution associated with clustering. Similar spectra are expected
for other extragalactic point sources, \eg radio galaxies.

\section{Phenomenological Theory}

{\bf Phenomenology:} We have progressed from the tens of pixels of
early $\Delta T/T$ experiments through the thousands for DMR and SK,
and will soon be dealing with tens of thousands for long duration
balloon experiments and eventually millions for the MAP and
COBRAS/SAMBA satellites. How best to analyze statistically these data
sets is a subject being developed largely by CMB theorists. Theorists
have also taken on an increasingly phenomenological role in LSS
studies and many of the same techniques are being applied to both
$\Delta T/T$ and LSS data sets. This is opening up into a major
subfield, a trend which we should strongly encourage. Finding nearly
optimal strategies for data projection, compression and analysis which
will allow us to disentangle the primary anisotropies from the
Galactic and extragalactic foregrounds and from the secondary
anisotropies induced by nonlinear effects will be the key to realizing
the theoretically-possible precision on cosmic parameters and so to
determine the winner (and losers) in theory space. Particularly
powerful is to combine results from different CMB experiments and
combine these with LSS and other observations.  Almost as important as
the end-product is the application of the same techniques to probing
the self-consistency and cross-consistency of experimental results.

\begin{figure}
\vspace{-.5in}
\centerline{\epsfxsize=8.0in\epsfbox{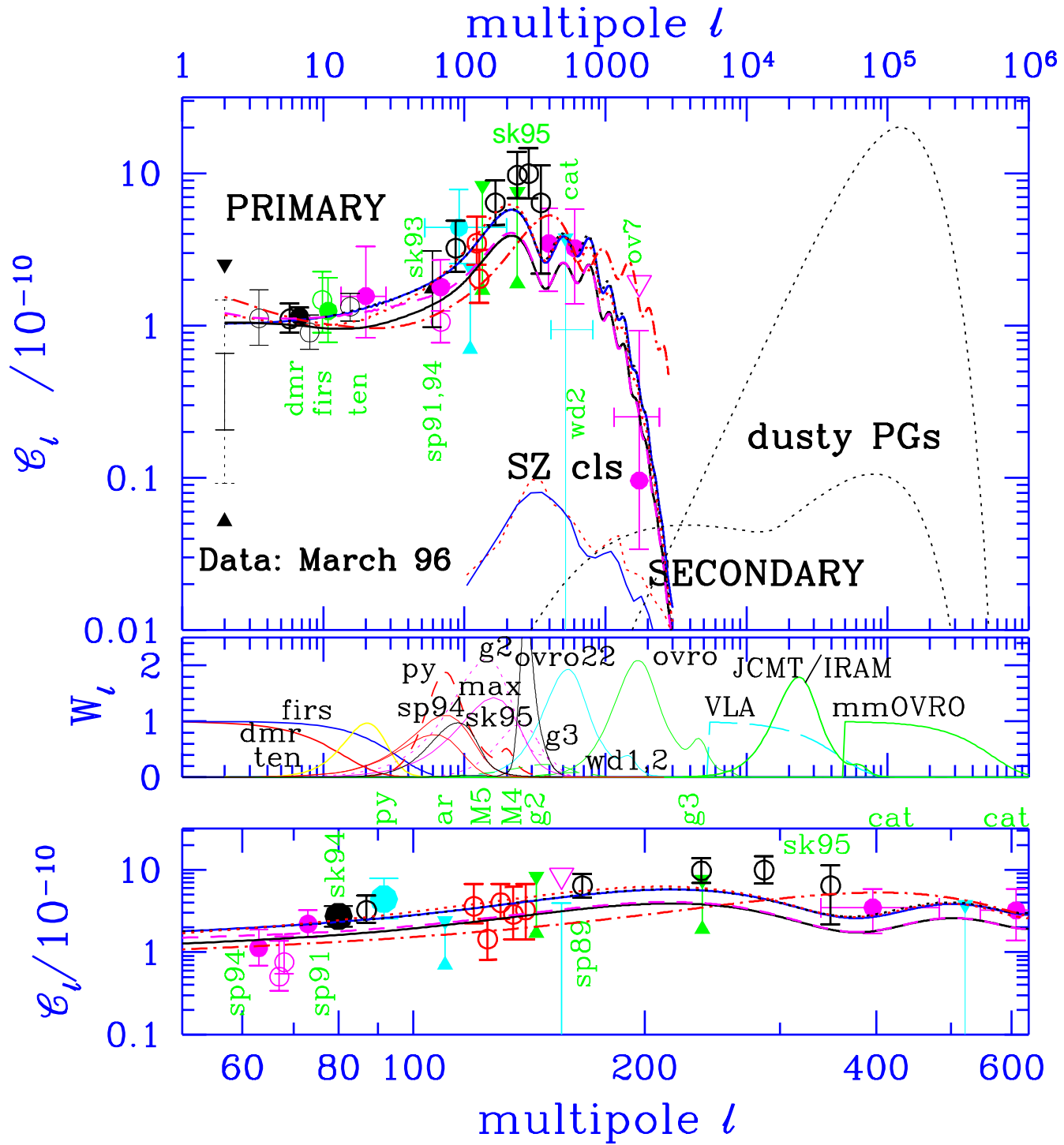}}
\vspace{-.3in}
\caption{Sample primary and secondary power spectra, ${\cal
C}_\ell \equiv \ell (\ell +1) \avrg{\vert (\Delta T/T)_{\ell m}\vert^2
}/(2\pi )$, are compared with the band-power estimates derived for the
anisotropy data up to March 1996. The lower panel is a closeup of the
first two `Doppler peaks'. Average filter functions for a variety of
experiments are shown in the middle panel.  The upper solid primary
${\cal C}_\ell$ curve is a COBE-normalized ``standard'' untilted CDM
model, and variants are shown with the same cosmological age and
$\Omega_B{\rm h}^2$, but nonzero tilt, $\Omega_{hdm}$ (\ie $m_\nu
>0$), $\Omega_{vac} $ (\ie $\Lambda > 0$), average curvature (\ie
$\Omega_{tot} < 1$) or weak reionization. They all broadly
agree with the data. By contrast, the thermal SZ anisotropy power is
way down, the kinematic SZ power is off-scale, and dusty emission
power from early galaxies is concentrated at higher $\ell$ and higher
frequency.}
\label{fig:CLpow}
\end{figure}

{\bf Current State:} Current band-powers, shown in
Fig.~\ref{fig:CLpow}, broadly follow inflation-based expectations, but
may still include residual signals. Consistency with the primary
anisotropy frequency spectrum has been shown for DMR and for the
smaller angle experiments, but over a limited range. That the level is
$\sim 10^{-5}$ provides strong support for the gravitational
instability theory. To get the large scale structure of
COBE-normalized fluctuations right provides encouraging support that
the initial fluctuation spectrum was not far off the scale invariant
form that inflation (and defect) models prefer. That there appears to
be power at $\ell \sim 400$ suggests the universe could not have
reionized too early.

\begin{figure}
\centerline{\epsfxsize=8.0in\epsfbox{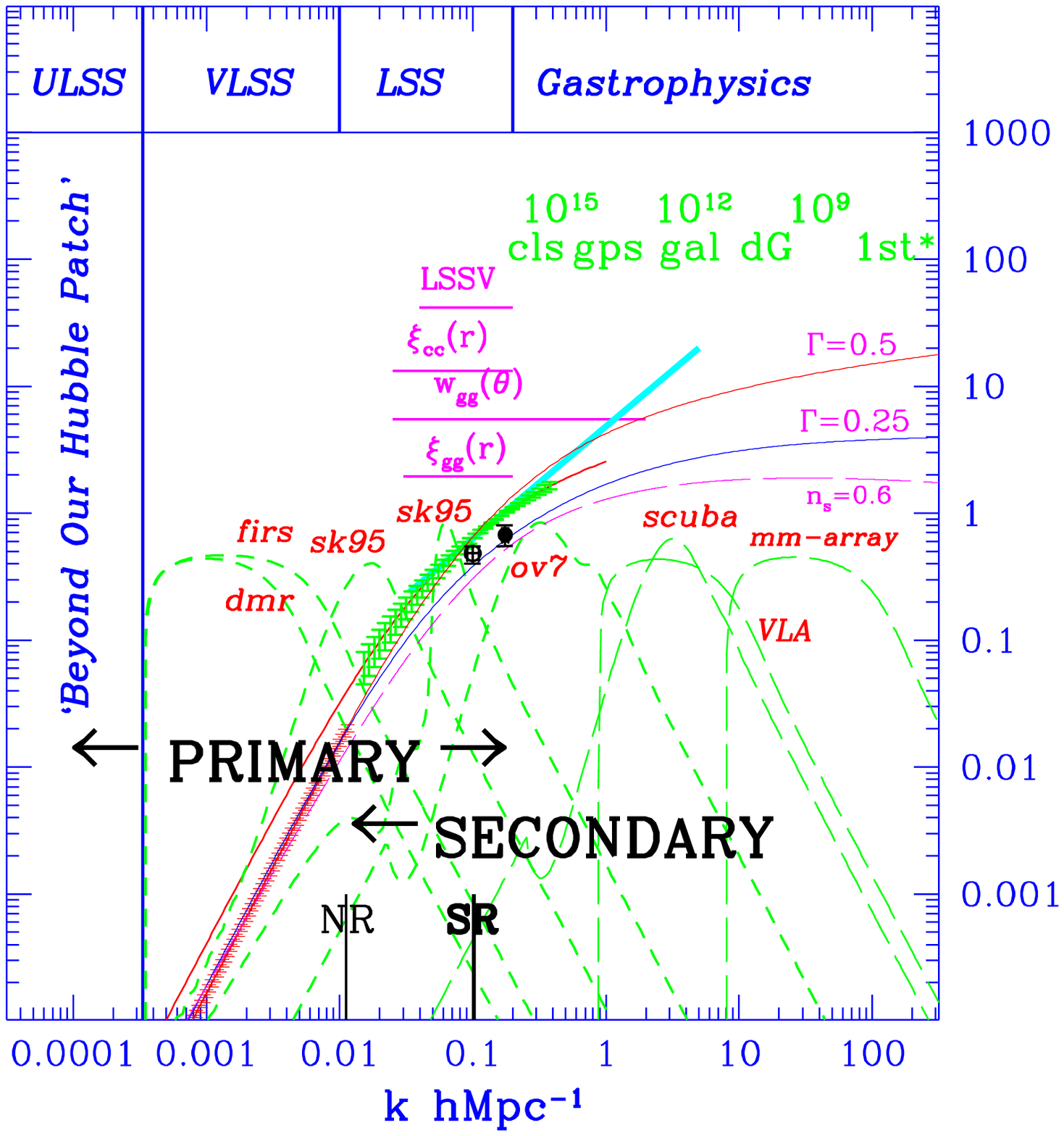}}
\vspace{-.2in}
\caption{The bands in comoving wavenumber probed by CMB primary and
secondary anisotropy experiments and by large scale structure
observations are contrasted. Sample (linear) density power spectra are
for the ``standard'' $n_s=1$ CDM model (labelled $\Gamma=0.5$), for a
tilted ($n_s=0.6$, $\Gamma=0.5$) CDM model and for a model with the
shape modified ($\Gamma=0.25$) by changing the matter content of the
Universe. A (uniform?) bias is allowed to raise the shapes into the
hatched $w_{gg}$ region; only the latter two fit. The solid data point
in the cluster-band denotes the constraint from the abundance of
clusters, and the open data point at $10 \hmpc$ a constraint from
 streaming velocities (for $\Omega_{tot}$=1,$\Omega_{vac}$=0).}
\label{fig:probes} 
\end{figure}

{\bf Large Scale Structure:} We have always combined CMB and LSS data
in our quest for viable models. Fig.~\ref{fig:probes} shows how the
two are connected. As we have seen, the DMR data precisely determines
$\sigma_8$ for each model considered. For the COBE-normalized density
power spectra to thread the ``eye of the needle'' associated with
cluster abundances severely constrains the parameters determining
them. Similar constrictions arise from galaxy-galaxy and
cluster-cluster clustering observations.  Smaller angle CMB data (\eg
SP94, SK95) are consistent with these models (\eg Bond and Jaffe,
these proceedings), and will soon be powerful enough for the CMB by
itself to offer strong selection, but this will definitely not
diminish the combined LSS-CMB phenomenology.

{\bf Ultra-large Scale Structure:} The ``beyond our horizon'' land in
Fig.~\ref{fig:probes} is actually partly accessible because long waves
contribute gentle gradients to our observables. Constraints on such
``global parameters'' as average curvature are an example, $d_c > 1.1
H_0^{-1}$, though not yet very restrictive. One may also probe whether
a huge bump or deficit in power exists just beyond $k^{-1} \sim
H_0^{-1}$, but this has not been much explored. The remarkable
non-Gaussian structure predicted by stochastic inflation theory would
likely be too far beyond our horizon for its influence to be felt. The
bubble boundary in hyperbolic inflation models may be closer and its
influence shortly after tunneling occurred could have observable
consequences for the CMB. Theorists have also constrained the scale of
topology in simple models; \eg we (Pogosyan, Sokolov and I) find
$d_T/2 > 2 H_0^{-1}$ for flat 3-tori and $> 1.5 H_0^{-1}$ for flat
1-tori from DMR (see de Oliveira-Costa, these proceedings). A number
of groups are now trying to constrain the compact hyperbolic
topologies.

{\bf Futures:} Let us look forward to the day phenomenological
theorists will have optimally-analyzed
LDBs/VSA/CBI/ChiSPI/MAP/COBRAS/SAMBA and we know the power spectrum
and cosmic parameters to wonderful precision. What will it mean? It
may not be clear. Take inflation as an example. There will be
attempts, undoubtedly optimal ones, to reconstruct the inflaton's
potential, but all of our CMB and LSS observations actually access
only a very small region of the potential surface, and even this will
be fuzzily determined if we allow too much freedom in parameter
space. Still even a fuzzy glimpse is worth the effort.  Most fun will
be when phenomenology teaches us that non-baroque inflation and defect
models fail and howling packs of theorists go hunting for the elusive
generator following trails well marked by the data.

\def\prd{{Phys.~Rev.~D}}
\def\prl{{Phys.~Rev.~Lett.}}
\def\apj{{Ap.~J.}}
\def\apjl{{Ap.~J.~Lett.}}
\def\apjsuppl{{Ap.~J.~Supp.}}
\def\mnras{{M.N.R.A.S.}}


\begin{table}[htbp]
  \begin{center}
    \leavevmode
    \begin{tabular}{|l|l|}
\hline
\multicolumn{2}{c}{PHYSICAL PROCESSES FOR ANISOTROPY} \\
\hline
\hline
\multicolumn{2}{c}{PRIMARY SCALAR ANISOTROPIES} \\
\hline
$\Phi /3$ & ``Naive'' Sachs-Wolfe effect: Gravitational Potential  \\
${1\over 4} {\delta \rho_\gamma \over \rho_\gamma }$ & photon
bunching (acoustic): ${1\over 3} {\delta \rho_B \over \rho_B }$ 
effect, isocurvature effect  \\
$\sigma_T \bar{n}_e \bv_e \cdot \hq$ & Linear-order Thompson
scattering (Doppler) \\
$2\int_{l.o.s.} \dot{\Phi}$ & Integrated Sachs-Wolfe effect  \\
  & subdominant anisotropic stress and polarization terms\\
\hline
\multicolumn{2}{c}{PRIMARY TENSOR ANISOTROPIES} \\
\hline
$\half \dot{h}_{+, \times}$ & gravity waves (two polarizations) \\
 & subdominant polarization terms\\
\hline
\multicolumn{2}{c}{SECONDARY ANISOTROPIES} \\
\hline
$\eps_{AB}$ & Linear Weak Lensing: 2D shear tensor \\
$2\int_{l.o.s.} \dot{\Phi}_{NL}$ & Rees-Sciama effect: Linear Response to $\Phi$ of nonlinear structure\\
$\sigma_T \delta {n}_e \bv_e \cdot \hq$ & Nonlinear Thompson
scattering: Quadratic-order (Vishniac) effect,  \\
$\ldots\ldots$    & ``kinematic'' SZ effect (moving cluster/galaxy) \\
$\int_{l.o.s.} \psi_K (x ) \delta (n_e T_e) $ &   thermal SZ
effect: 
Compton cooling from nonlinear
gas ($x=E_\gamma/T_\gamma$)
\\
$ \int_{l.o.s.} \psi_{dust}(x_d)\rho_d$ & redshifted dust emission,
pregalactic/protogalactic ($x_d=E_\gamma/T_d$)
\\
\hline
\multicolumn{2}{c}{FOREGROUNDS} \\
\hline
& extragalactic radio sources: falling, flat, rising \\
&  IRAS sources and extrapolations to moderate $z$ \\
 &      Galactic bremsstrahlung, synchrotron \\
 &      Galactic Dust: regular, cool, strange?? \\
\hline
\end{tabular}
    \label{tab:process}
  \end{center}
\end{table}

\begin{thebibliography}{99}{\baselineskip 0.1cm
\bibitem{bh95} Bond, J.R. 1996, {\it Theory and Observations of the
Cosmic Background Radiation}, in ``Cosmology and Large Scale
Structure'', Les Houches Session LX, August 1993, ed. R. Schaeffer,
Elsevier Science Press, and references therein.  

\bibitem{comba95}
Bertschinger, E., Bode, P., Bond, J.R., Coulson, D., Crittenden, R.,  
Dodelson, S., Efstathiou, G., Gorski, K., Hu, W.,  Knox, L., Lithwick,
Y., Scott, D., Seljak, U., Stebbins, A., Steinhardt, P., Stompor, R., 
Souradeep, T., Sugiyama, N., Turok, N., Vittorio,  N., White,  M.,
Zaldarriaga, M.  1995, 
ITP workshop on {\it Cosmic Radiation Backgrounds and
the Formation of Galaxies}, Santa
Barbara.  


}\end{thebibliography}
\end{document}